\documentstyle[11pt,aaspp4,tighten]{article}

\newcommand{\lsun}{log $L/L_{\odot}$} 
\newcommand{\msun}{$M/M_{\odot}$}
\newcommand{\px}{$P_1 / P_0$~}
\newcommand{\pa}{$P_0$~}
\newcommand{\pb}{$P_1$~}

\begin{document}

\title{DOUBLE-MODE RR LYRAE VARIABLES: PULSATIONAL MASSES REVISITED}

\author{Giuseppe Bono}
\affil{Osservatorio Astronomico di Trieste, Via G.B. Tiepolo 11,
34131 Trieste, Italy; bono@oat.ts.astro.it}

\author{Filippina Caputo}
\affil{Osservatorio Astronomico di Capodimonte, Via Moiariello 16,
80131 Napoli, Italy; caputo@astrna.na.astro.it}

\author{Vittorio Castellani}
\affil{Dipartimento di Fisica, Univ. di Pisa, Piazza Torricelli 2,
56100 Pisa, Italy; vittorio@astr1pi.difi.unipi.it}

\and
\author{Marcella Marconi}
\affil{Dipartimento di Fisica, Univ. di Pisa, Piazza Torricelli 2,
56100 Pisa, Italy; marcella@astr1pi.difi.unipi.it}

\pagebreak
\begin{abstract}

\noindent 
Double-mode RR Lyrae variables (i.e. radial variables which are
simultaneously pulsating in both fundamental and first overtone
modes) appear a fundamental tool for investigating the mass of old 
Population II Horizontal Branch (HB) stars. The most widespread method
adopted for evaluating the masses of these objects is based on the 
Petersen (1973) approach, which relies only on pulsational periods
(\px vs. \pa) and therefore is independent of any preliminary evaluation 
of the reddening and/or of the distance modulus of the stellar cluster. 

\noindent
In this paper we
supply an overview of the mass estimates and underline the role played
by opacities as well as by full amplitude nonlinear models for
removing the discrepancy between pulsational and evolutionary masses.
On the basis of the comparison between the theoretical scenario and 
double-mode RR Lyrae stars belonging to selected Galactic globular 
clusters (IC4499, M3, M15, M68, NGC2419, NGC6426) we show that 
the \px vs. \pa diagram can provide valuable constraints also on the 
luminosity of these variables.
\end{abstract}

\noindent
{\em Subject headings:} globular clusters: individual (IC4499, M3, M15, 
M68, NGC2419, NGC6426) -- stars: horizontal branch -- stars: oscillations 
-- stars: variables: other 

\pagebreak
\section{INTRODUCTION}

\noindent 
The evolution and the final fate of stellar structures is mainly
governed by the amount of original mass. According to such a
plain evidence, the large amount of present theoretical interpretations
of the evolutionary status of stars and stellar systems is  providing
tight constrains on the mass of the investigated objects. An independent
estimation of stellar masses would be of course of paramount interest
since it would represent a {\em prima facie} evidence for the physical 
reliability of the adopted evolutionary scenario. As it is 
well known, radial pulsating
structures offer such an opportunity for the simple reason that the
pulsations are mainly governed by gravity. On this simple basis,
the periods should depend, as they actually do,
on pulsator masses and radii (i.e. on the stellar parameters M, L and Te).
Jorgensen \& Petersen (1967) originally suggested that the occurrence
of double-mode pulsators could give the unique opportunity to provide
a straightforward evaluation of pulsator masses taking into account only
the ratio between the fundamental (\pa) and the first overtone
periods (\pb). 

\noindent 
According to this scenario, Petersen (1973) 
introduced the diagram \px vs. \pa (hereinafter referred to as PD) 
as a suitable tool for estimating the actual  mass value of double-mode
pulsators. The application of this procedure to
RR Lyrae stars dates back to Cox, King \& Hodson (1980). 
On the basis of the PD they found a mass value of about \msun=0.65 for
the only double-mode RR Lyrae (RRd) known at that time (AQ Leonis,
Jerzykiewicz \& Wenzel 1977). Since then the discovery of new RRd 
variables in several Galactic globular clusters, in the field 
(Clement et al. 1986, hereinafter referred to as C86)
and in dwarf spheroidal galaxies (Nemec 1985a, Kaluzny et al. 1995) has 
brought on an interesting discussion for constraining their
pulsational and evolutionary characteristics. Cox, Hodson
\& Clancy (1983, hereinafter referred to as CHC) 
investigated the PD for RRd pulsators in the metal poor Oosterhoff II
cluster M15 and  derived a pulsational mass of the order of 
M/${M_{\odot}}$ =0.65. The same authors suggested a mass of the order 
of \msun= 0.55 for the two RRd pulsators belonging to M3, the prototype
of intermediate metallicity Oosterhoff type I (Oo I) clusters.
These results were subsequently confirmed by Nemec (1985b) and by
C86 who found similar mass values for RRd variables in the Oo 
I cluster IC4499.

\noindent 
However, for the quoted HB pulsators current evolutionary theories foresee 
larger masses, namely \msun$\simeq$0.8 and $\simeq$0.65 for
Oo II and Oo I clusters respectively (see Bono et al. 1996). 
Such a disturbing discrepancy between the masses of RRd variables determined 
from pulsational and from evolutionary theories was settled as soon as 
Cox (1991) found that pulsational models incorporating new and updated
opacity evaluations were able to reconcile pulsational
and evolutionary predictions. The settling of this long-standing
discrepancy was thus regarded as an evidence
for  the reliability of the new opacity tables. 

\noindent 
In this paper we 
present a new investigation of the Petersen approach which
discloses some unexpected results and sheds  new light on the matter.
We show that opacity, as originally suggested by Cox (1991,1995), is the  
key physical ingredient which produces the disagreement between
pulsational and evolutionary masses. However, we also find that this 
discrepancy, even by using old opacities, can be consistently removed 
either by adopting a much finer spatial resolution in linear computations 
or by relying on detailed nonlinear models. Nevertheless, an exhaustive 
solution to the problem of RR Lyrae masses can only be achieved if both 
new opacities and full amplitude, detailed, nonlinear, nonlocal and 
time-dependent convective models are taken into account. 

\section{MASSES AND LUMINOSITIES OF RRd VARIABLES}   

\noindent 
During the last few years we have carried out an extensive survey of
limiting amplitude, nonlinear models of RR Lyrae variables (Bono
\& Stellingwerf 1994, hereinafter referred to as BS). The main purpose 
of this project is to examine 
the dependence of modal stability and pulsation behavior on astrophysical
parameters (for complete details see Bono et al. 1996). As a by-product of 
this investigation we revisited the problem of 
pulsator masses by investigating the dependence of the Petersen
diagram on the various assumptions governing theoretical calculations.

\noindent 
The sequences of static envelope models were analyzed in the linear 
nonadiabatic approximation (Castor 1971) and each model was required to
cover the outer 90\% of the stellar photospheric radii. The outer boundary 
condition was typically fixed at an optical depth of the order of 0.001.
The linear models were constructed by neglecting convection and by adopting 
the analytical approximation of old Los Alamos "King" opacity tables 
provided by Stellingwerf (1975a,b). On the basis of these assumptions 
a typical {\em coarse} model is characterized by 100-150 zones and few 
percents of the total stellar mass. Complete details of the mass ratio
between consecutive zones and the method adopted for constraining the 
hydrogen ionization region are given in Stellingwerf (1975a) and BS.  

\noindent 
As a starting point, Fig. 1 shows the theoretical PD obtained for 
selected values of stellar masses and luminosities. The models plotted 
in this figure present a stable {\em linear} limit cycle in the first 
two modes.  To understand the meaning of theoretical data displayed in 
this figure, we recall
that linear models provide evaluations of periods independently of the 
actual limit cycle stability of a given mode. As a consequence, we have 
to bear in mind that a rather large
amount of data in similar figures should be regarded as unphysical,
since they supply the ratio \px even where either the fundamental or
the first overtone modes present an unstable {\em nonlinear} limit cycle.

\noindent 
The period ratios of M15 RRd variables plotted in Fig. 1 were evaluated 
taking into account different estimates (Nemec 1985b; Kovacs, Shlosman 
\& Buchler 1986; Clement \& Walker 1990; Purdue et al. 1995). 
The error bar plotted in the lower right corner is referred to these 
measurements.
The comparison between theoretical models and observational data, 
shown in the above figure, clearly supports previous results given in 
the literature under similar theoretical assumptions and discloses  
the occurrence of the "mass discrepancy problem". At the same time, 
Fig. 1 shows 
that at a given fundamental period the period ratio \px appears largely 
independent of the assumed luminosity level.

\noindent 
In order to investigate the dependence of linear periods on the spatial 
resolution previously adopted, a new set of linear {\em detailed} models 
have been 
computed by adopting  the prescriptions suggested by BS. The number of 
zones for these new sequences of models is increased by roughly a factor 
of two with respect to the {\em coarse} ones and ranges from 200 to 300.
Fig. 2 shows the results of these new computations, disclosing that 
the "mass discrepancy problem" appears affected also by the method adopted
to discretize the physical structure of the static envelope model. 
As a matter of fact, we find that periods provided by linear, nonadiabatic,  
radiative models constructed with a finer spatial resolution partially
remove the degeneracy of the luminosity levels. Moreover, as a most 
relevant point, these calculations now suggest that the mass value of 
Oo II RRd variables should be of the order of \msun= 0.8, whereas
Oo I  RRd variables should increase to about \msun= 0.70, in much 
better agreement with evolutionary prescriptions (see Bono et al. 1996). 

\noindent 
However, BS have already shown that linear
periods are only a first, though good approximation of the
pulsational periods obtained from a more appropriate
nonlinear treatment of the pulsation. Thus the problem arises if
linear predictions about RRd masses are preserved in the nonlinear 
approach. To properly address this fundamental theoretical question,  
Fig. 3 displays the results of several sequences of nonlinear, nonlocal 
and time-dependent convective models constructed by assuming the same 
equation of state and the same opacities adopted in the linear regime. 
According to the negligible influence of 
spatial resolution on nonlinear limiting amplitude characteristics and 
modal stability (BS and references therein) in order to speed up the 
calculations required by the nonlinear approach only {\em coarse} static 
envelope models were taken into account.

\noindent 
The dynamical behavior of the envelope models was examined for the first
two modes and the static structures were forced out of equilibrium by 
perturbing the linear radial eigenfunctions with a constant velocity 
amplitude of 20 km$s^{-1}$. The method adopted for initiating nonlinear
models unavoidably introduces a spurious component of both periodic 
and nonperiodic fluctuations which are superimposed to the pure radial 
motions. As a consequence, before the dynamical behavior approaches 
the limit cycle stability it is necessary to carry out extensive 
calculations. The fundamental and first overtone sequences have been 
integrated in time for at least 2,000 periods. The models located close 
to the fundamental blue edge and to the first overtone red edge were 
followed for a longer time interval (2,000-6,000 periods) since in these 
regions of the instability strip before the dynamical motions approach 
their asymptotic behavior a switch-over to a different mode could 
take place even after several thousand periods. 
The integration is generally stopped as soon as the nonlinear work 
term is vanishing and the pulsational amplitudes present a periodic
similarity of the order of $10^{-(4 \div 5)}$. 

\noindent 
Therefore it turns out that the decrease of theoretical points plotted in 
Fig. 3 is tightly connected with the morphology of the "OR region", since
were taken into account only envelope models which present stable nonlinear 
limit cycles both in the fundamental and in the first overtone modes. 
Moreover, data in Fig. 3 reveal that the nonlinear PD differs intrinsically 
from the canonical linear PD. In fact in this new context the
spurious theoretical points connected with models which present a
unique stable limit cycle (fundamental or first overtone) 
have obviously disappeared. A direct interesting consequence of this 
new theoretical scenario is that the comparison between nonlinear 
periods and observational data can now give useful information on both 
stellar masses and luminosities of the pulsators. The reader interested 
to a thorough analysis concerning 
the evaluation of these parameters on the basis of RRd variables belonging 
to both Oo I and Oo II clusters is also referred to Cox (1995) and 
Walker (1995). In the evaluation of masses we eventually find that
nonlinear results do not fully support linear indications. 
In fact, on the basis of nonlinear periods we obtain a stellar mass
of \msun$\simeq$0.7 for Oo II cluster  pulsators, whereas for 
the RRd variables in IC4499 we estimate a mass of the order of 
\msun=0.60.
As a consequence, the agreement found by relying on linear detailed models
has to be regarded as an artifact of the computational procedure.
Moreover, for M15 and M68 pulsators we find a luminosity
around \lsun$\simeq$1.8, which appears somewhat larger than the  
currently accepted evolutionary predictions.

\noindent 
Bearing in mind the present scenario, we now take into account the 
effects of the new opacities  provided by Rogers \& Iglesias (1992) 
for temperatures higher than $10^4$ $^oK$ and by Alexander \& Ferguson 
(1994) for lower temperatures. The reader interested in the method 
adopted for handling the new opacity tables is referred to Bono, Incerpi 
\& Marconi (1996). 
For the sake of conciseness, 
we briefly  quote the mass evaluations obtained from linear computations:
\msun= 0.72, 0.60  (coarse models) and \msun= 0.78, 0.65 (detailed models) 
for pulsators in Oo II and Oo I clusters respectively. 
Fig. 4 shows nonlinear periods based on updated radiative opacities.
The comparison with observational data is now pointing out a promising
theoretical scenario since it predicts a stellar mass slightly greater 
than \msun= 0.8 for RRd pulsators in Oo II clusters and a mass value 
around \msun= 0.65  for RRd variables in IC4499. 
Both results are now in excellent agreement, within the error bar, with 
canonical evolutionary predictions. 
Data plotted in Fig. 4 also suggest a luminosity level of the order 
of \lsun $\simeq$1.7 for Oo II RRd variables, whereas the corresponding
luminosity level for RRd variables in IC4499 falls between the computed
luminosity levels at \lsun =1.61 and 1.72. The overall good agreement  
with evolutionary predictions, presented in Bono et al. (1996), shows
that thanks to the updated physical input both pulsational and evolutionary
theories converge to form a homogeneous scenario concerning
the long debated question of RR Lyrae luminosity in globular clusters. 

\noindent 
Finally, it is worth noting that the two RRd variables in M3 appear 
slightly more massive and more luminous than RRd variables in IC4499. 
According to current metallicity estimates for these clusters 
($[Fe/H]_{M3}$=-1.7, $[Fe/H]_{IC4499}$=-1.5), even this finding appears 
again in satisfactory agreement with the evolutionary prescriptions.

\section{CONCLUSIONS} 

\noindent 
In this paper we have revisited the approach based on the PD
for determining the masses of RRd variables. It is shown that the 
pulsator masses evaluated through the comparison between periods obtained 
in a linear, nonadiabatic, radiative regime and observational data might 
be affected by substantial systematic errors. On the other hand, the periods 
provided by the surveys of nonlinear, nonlocal and time-dependent convective 
models point out that even though the discrepancy between linear and 
nonlinear periods has often been considered negligible, it plays a key role 
for properly defining the location of double-mode pulsators inside the 
Petersen diagram (\px vs. \pa).

\noindent 
As a most relevant point, 
we found that a nonlinear Petersen diagram constructed taking
simultaneously into account both nonlinear models and new
radiative opacities provides valuable constraints not only
on the stellar masses but also on the luminosities of RRd
variables. The pulsational masses and luminosities of
double-mode pulsators obtained in this new theoretical framework confirm 
the results recently provided by Cox (1995). 
The comparison with observational data of RRd variables in both Oo I and
Oo II galactic globular clusters discloses a satisfactory agreement 
with current evolutionary and pulsational predictions. At the same time,
this agreement supplies a new piece of evidence against the suggested 
anomaly of HB star luminosities. 
Further applications of this new approach for constraining the physical 
parameters of RRd variables belonging to the Galactic field, the central
region of LMC and to dwarf spheroidal galaxies are under way. 

\noindent 
It is a pleasure to thank A. Cox as referee for several valuable comments 
and for the pertinence of his suggestions on the original version of this 
paper. This work was partially supported by MURST, CNR-GNA and ASI. 
\pagebreak

\pagebreak
\section{Figure Captions}

\vspace*{3mm} \noindent {\bf Fig. 1.} - 
Petersen diagram for linear, nonadiabatic and radiative periods. The 
RR Lyrae models were computed by adopting a fixed chemical composition 
(Y=0.24, Z=0.0001) and the labeled values of stellar masses and 
luminosity levels. These envelope models were constructed by adopting
the analytical opacity approximation provided by Stellingwerf (1975a,b).
RRd variables observed in Oo I clusters are marked by 
{\em open squares} (IC4499, C86) and by {\em stars} (M3, CHC), whereas 
those observed in the Oo II clusters are displayed by {\em open triangles}
(M68, Walker 1994), {\em cross} (NGC6426, Clement \& Nemec 1990) and 
{\em diamond} (NGC2419, Clement \& Nemec 1990). The period ratios of 
RRd variables belonging to M15 ({\em open circles}) were evaluated 
taking into account the estimates provided by Nemec (1985b), Kovacs, 
Shlosman \& Buchler (1986), Clement \& Walker (1990) and 
Purdue et al. (1995). The error bar 
plotted in the lower right corner is referred to these measurements. 

\vspace*{3mm} \noindent {\bf Fig. 2.} - 
Same as Fig. 1, but theoretical periods are referred to linear RR Lyrae 
models characterized by a finer spatial resolution. 

\vspace*{3mm} \noindent {\bf Fig. 3.} - 
Petersen diagram for limiting amplitude, nonlinear, nonlocal and 
time-dependent convective models. Only models which present a stable 
nonlinear limit cycle both in the fundamental and in the first overtone 
are displayed. These models were computed, for the labeled values of 
stellar masses and luminosity levels, by adopting the Stellingwerf 
analytical approximation and a coarse spatial resolution. 

\vspace*{3mm} \noindent {\bf Fig. 4.} - 
Same as Fig. 3, but theoretical periods are referred to nonlinear 
RR Lyrae models computed by adopting the new opacity tables provided by 
Rogers \& Iglesias (1992) for temperatures higher than $10^4$ K, 
and by Alexander \& Ferguson (1994) for lower temperatures. 
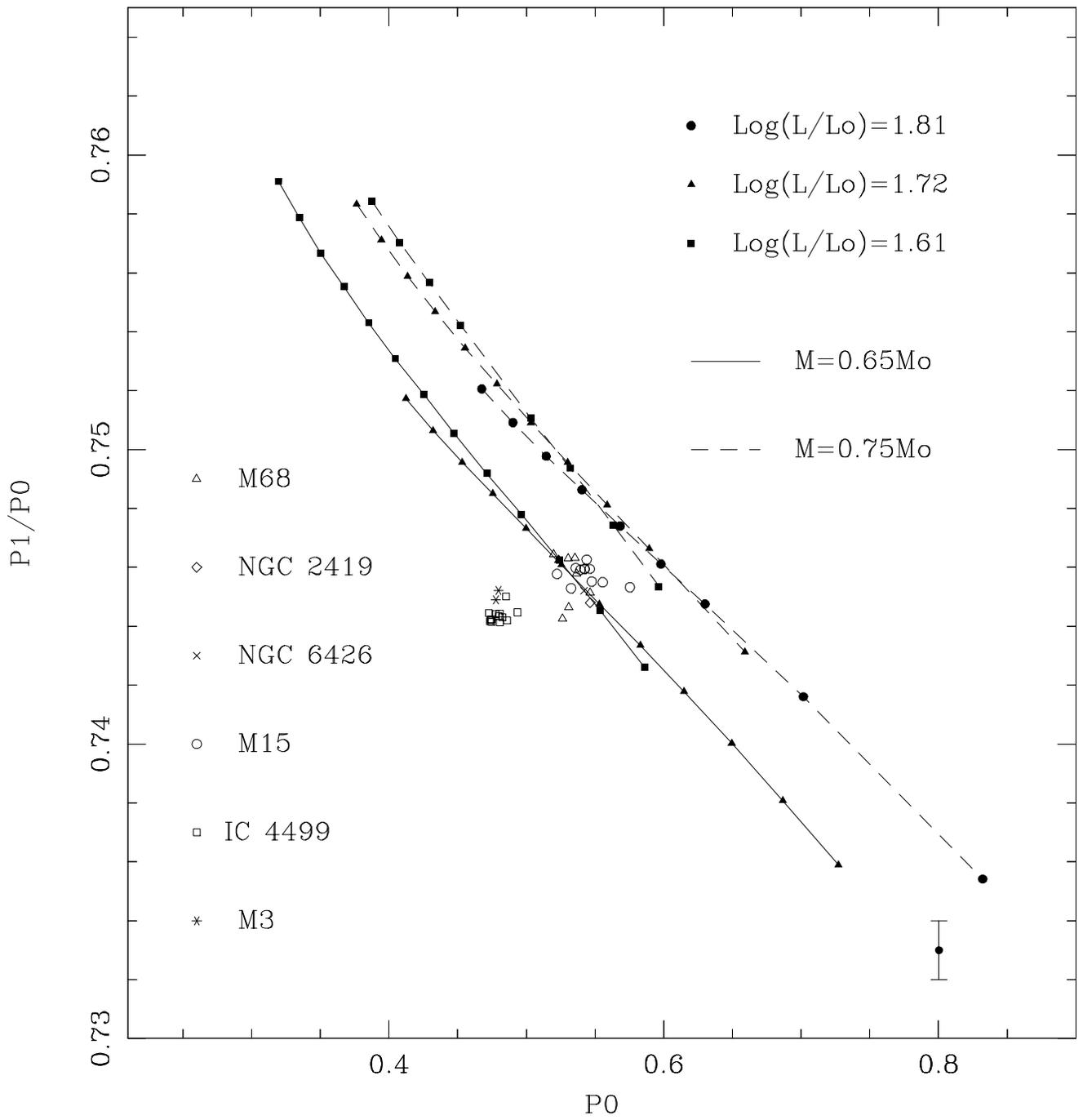
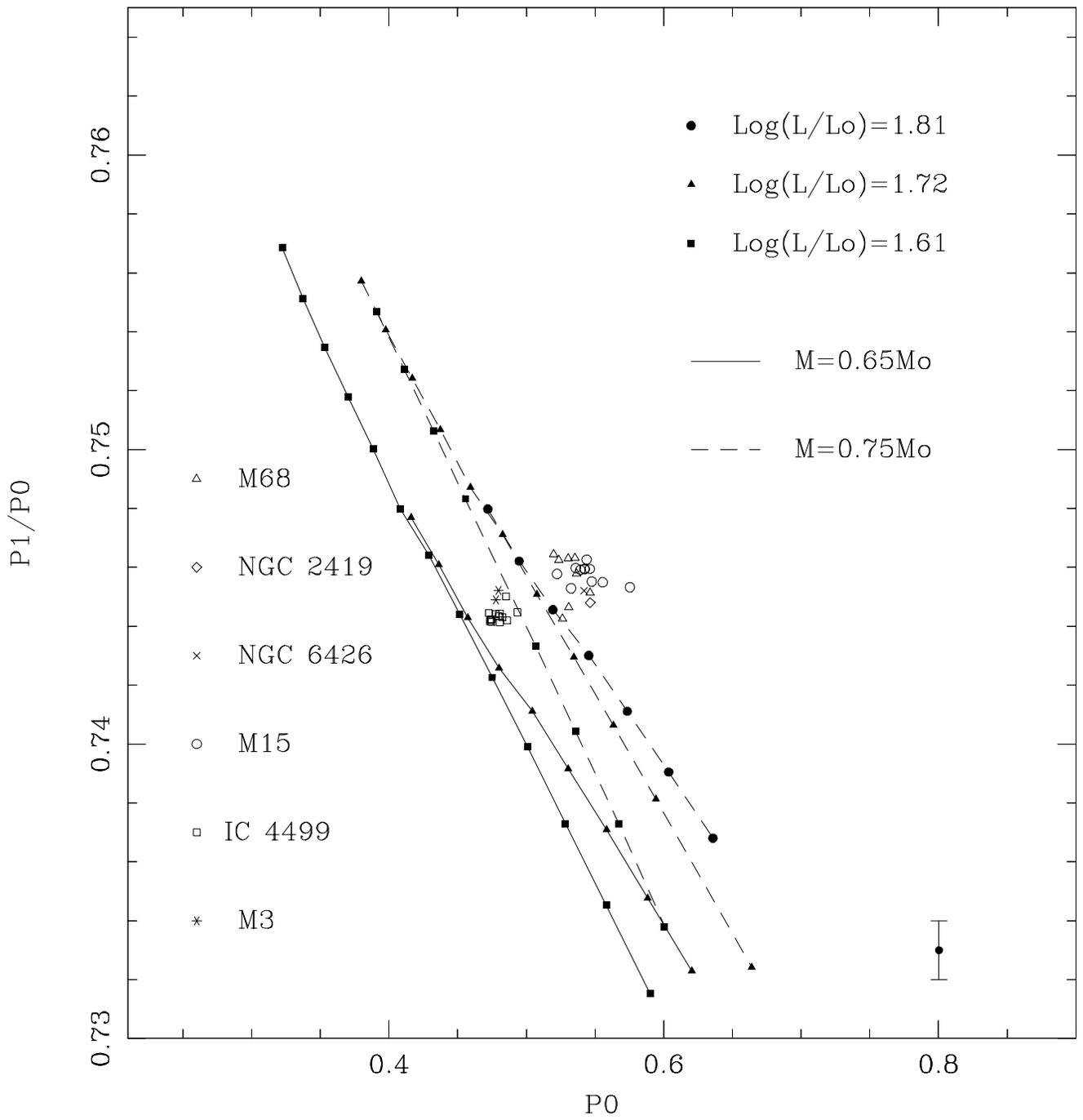
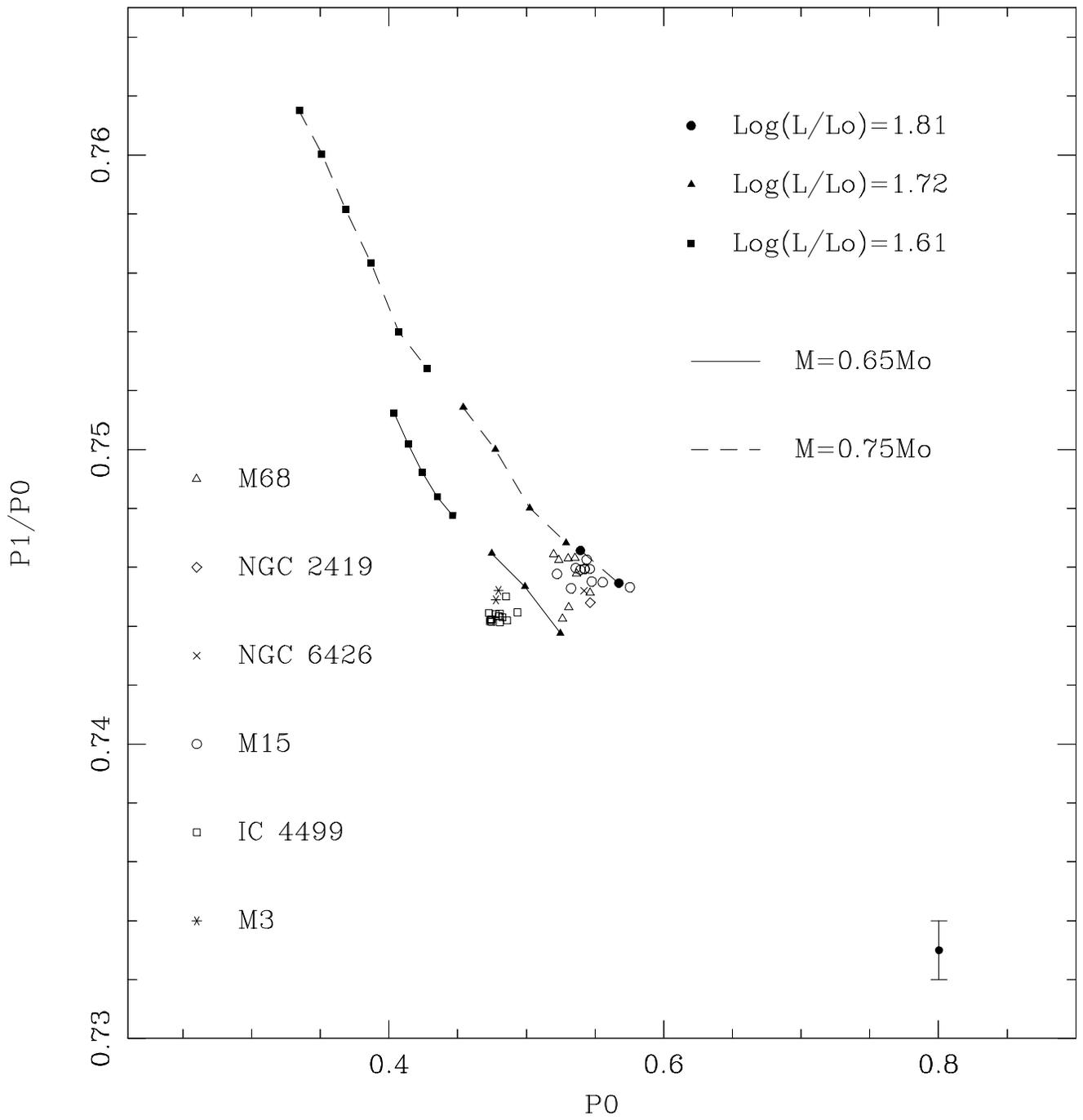
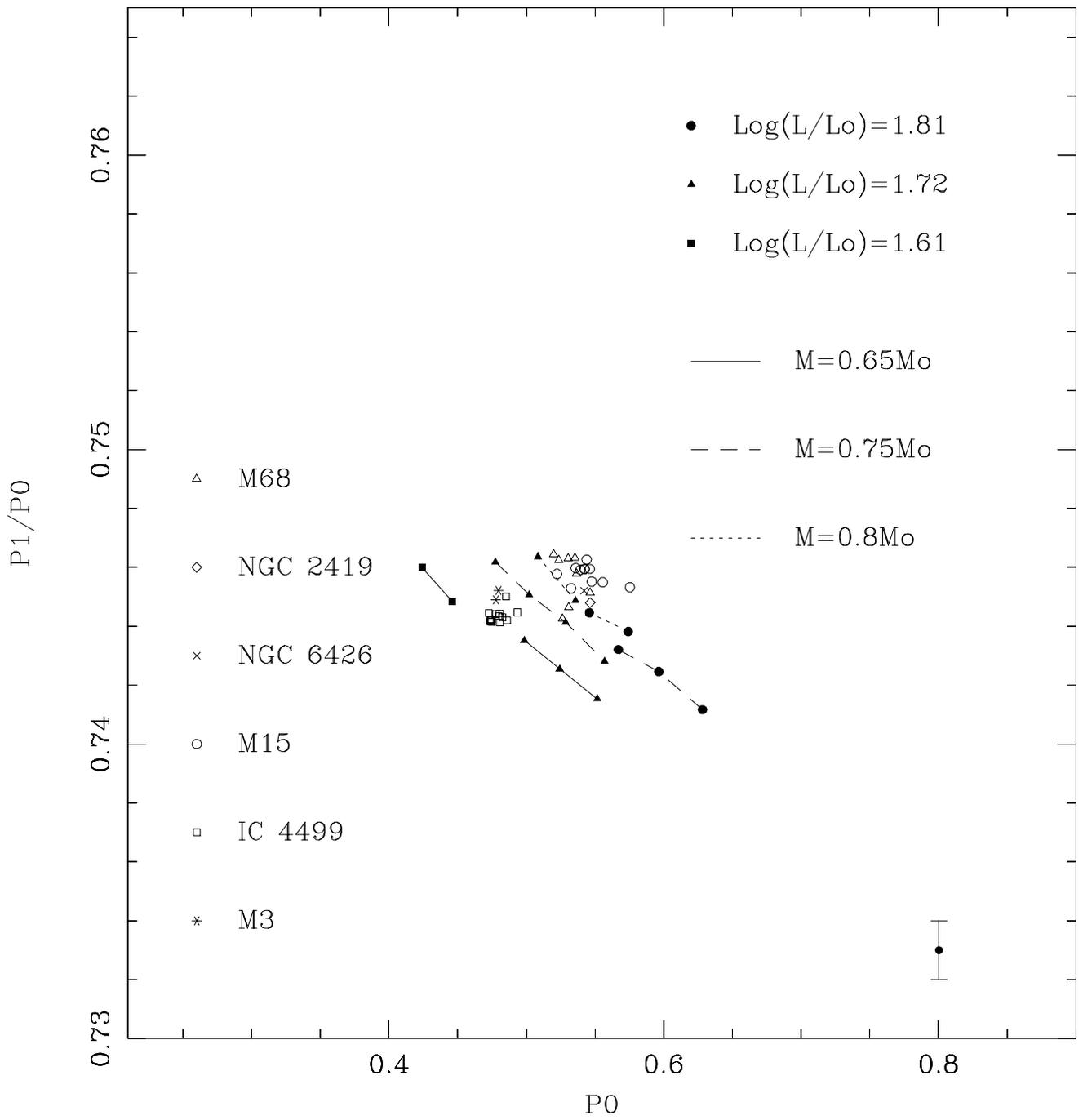
\end{document}